\documentclass{PoS}

\title{Recent results from HADES on electron pair production in relativistic heavy-ion collisions}
\ShortTitle{Probing resonance matter with virtual photons}

\author{\speaker{Tetyana Galatyuk} for the HADES Collaboration\\
        Goethe-Universit\"at Frankfurt am Main, Germany\\
        E-mail: \email{t.galatyuk@gsi.de}}
\newcommand{\ee}{$e^{+}e^{-}$}
\newcommand{\mee}{M$_{e^{+}e^{-}}$}

\newcommand{\xcc}{$^{12}C$+$^{12}C$}
\newcommand{\caca}{$^{40}Ca$+$^{40}Ca$}
\newcommand{\arkcl}{$^{40}Ar$+$^{39}KCl$}
\newcommand{\np}{$n$+$p$}
\newcommand{\pp}{$p$+$p$}
\newcommand{\pd}{$p$+$d$}
\newcommand{\xdp}{$d$+$p$}
\newcommand{\gevu}{GeV/$u$}
\newcommand{\gevcc}{GeV/$c^{2}$}

\newcommand{\gev}{GeV}
\newcommand{\mev}{MeV}
\newcommand{\etal}{$et~al.$}
\abstract{Systematic investigations of dilepton production are performed at the SIS accelerator of GSI with the HADES spectrometer. The goal of this program is a detailed understanding of di-electron emission from hadronic systems at moderate temperatures and densities. New results obtained in HADES experiments focussing on electron pair production in elementary collisions are reported here. They pave the way to a better understanding of the origin of the so-called excess pairs earlier on observed in heavy-ion collisions by the DLS collaboration and lately confirmed in two measurements of the HADES collaboration using \xcc\ and \arkcl\ collisions. Results of these studies are discussed.}
\FullConference{5th International Workshop on Critical Point and Onset of Deconfinement - CPOD 2009,\\
		 June 08 - 12 2009\\
		 Brookhaven National Laboratory, Long Island, New York, USA}
\begin{document}
%==============================================================================

%==============================================================================
\section{Dileptons as a probe of extreme matter}
%==============================================================================
The study of the electromagnetic structure of hadrons plays an important role in understanding the nature of matter. In particular the emission of lepton pairs out of the hot and dense collision zone in heavy-ion reactions is a promising probe to investigate in-medium properties of hadrons and in general the properties of matter under such extreme conditions. One challenge is to detect new phases of matter in the laboratory by isolating unambiguous signals. Among the promising observables are short-lived vector mesons decaying into lepton pairs inside the hot and dense medium. Such purely leptonic final states carry important information of the decaying objects to the detectors without being affected by strong final-state interaction while traversing the medium. Dileptons, however, are a very rare probe. While the production of real photons is suppressed relative to hadrons by about one power of the electromagnetic coupling constant, $\alpha = 1/137$, dilepton emission (i.e.\, virtual photons with subsequent decay $\gamma^{*} \rightarrow e^{+}e^{-}$ or $\gamma^{*} \rightarrow \mu^{+}\mu^{-}$) is further suppressed by an additional power of $\alpha$. The branching ratios for hadronic decays of vector mesons are thus typically $4$~orders of magnitude larger, unless they are suppressed by phase space factors. This is a severe penalty factor compared to hadronic decay channels like e.g.\ the $\rho \rightarrow \pi\pi$ decay. However, the strong final-state interactions of pions in a dense medium make an interpretation of the signal at least model-dependent, if not impossible. Observed effects could be interpreted both as due to medium modifications of the vector meson spectral function or due to the final-state interaction of the decay products.

Dilepton spectra measured by the CERES~\cite{ceres_pb_au_158_7per_centrality} and NA60~\cite{na60_prl} experiments at CERN-SPS energies ($40 - 158$~\gevu) demonstrated a significant in-medium modification of the $\rho$~meson spectral function signaled by an additional yield (excess) of lepton pairs in the invariant-mass region below the $\rho$~meson pole mass. The yield and spectral shapes of the CERES/NA60 results are well described by a broadening scenario for the $\rho$, but are not consistent with a dropping mass scenario (discussed e.g.\ in Ref.~\cite{ceres_pb_au_158_7per_centrality}). The broadening-mass scenario implies a strong coupling of the $\rho$ meson to baryons which adds strength to the dilepton yield at low invariant masses.

At beam energies of $1 - 2$~\gevu~~di-electron production was studied by the DLS\footnote{DiLepton Spectrometer} collaboration at the Bevalac~\cite{dls_prl_porter}. A large di-electron excess over the "hadronic cocktail" was observed in \xcc\ and \caca\ collisions. However, in contrast to the high-energy experiments, for a long time the excess could not be explained by any theoretical model and the situation hence became famous as the "DLS puzzle". In the SPS energy regime, the produced pions outnumber the existing baryons, whereas at SIS/Bevalac energies the pions hide themselves by forming excited baryon states ($\Delta$, $N^*$) which carry the produced heat to be released in subsequent collisions. Finally, at a late stage, the pion-depleted system releases the pions via decay of the resonances. This scenario obviously requires a different theoretical treatment than the pion-dominated high-energy regime.

\begin{figure}[tb]
\centering
\begin{minipage}[c]{0.5\textwidth}
  \includegraphics[width=0.9\textwidth]{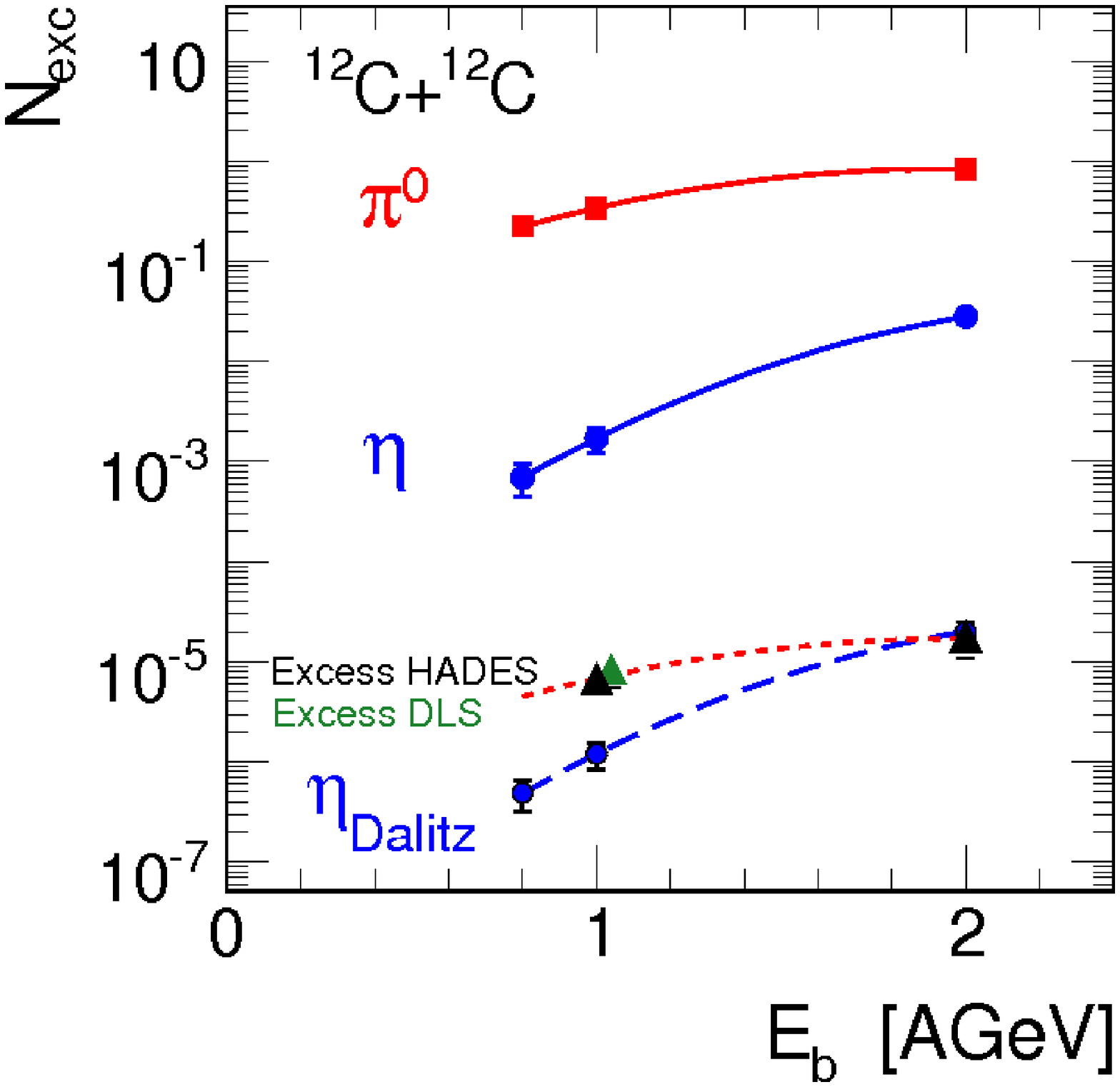}
\end{minipage}
\begin{minipage}[c]{0.45\textwidth}
  \caption{Inclusive multiplicity of the pair excess (N$_{exc}$) in the mass range
           \mee$=0.15 - 0.50$~\gevcc\ as a function of a beam energy E$_{b}$ (black
           triangles: HADES, green triangle: DLS)~\cite{hades_cc1gev,hades_cc2gev}.
           Also shown are the $\pi^{0}$ (red circles) and $\eta$ (blue circles,
           blue line) inclusive multiplicities
           in \xcc\ collisions~\cite{taps}, as well as the corresponding $\eta$
           Dalitz decay (black circles, blue dashed line) contribution integrated
           over \mee~$=0.15 - 0.50$~\gevcc. For comparison with N$_{exc}$ the
           down-scaled $\pi^{0}$ curve is shown as dashed red line. See~\cite{hades_cc1gev} for details.}
   \label{hades_excess}
\end{minipage}
\end{figure}

The excess of electron pairs in \xcc\ collisions was recently re-investigated by the HADES\footnote{High Acceptance Di-Electron Spectrometer} experiment at the GSI Helmholtzzentrum f{\"u}r Schwerionenforschung (Germany) for beam energies of $1$ and $2$~\gevu~\cite{hades_cc1gev,hades_cc2gev}. The excitation function of the excess-pair multiplicity with masses larger than those from $\pi^0$ Dalitz decays, i.e.\ in the mass range from $0.15$~\gevcc\ to $0.5$~\gevcc, are shown in Fig.~\ref{hades_excess} together with the pair multiplicity from $\eta$ Dalitz-decays (BR$_{{\gamma} e^{+}e^{-}}=0.6\%$) within the same  mass range. Within experimental errors, the excess scales like the $\pi^{0}$ multiplicity, not like the $\eta$ multiplicity, as is demonstrated by the direct comparison of the excess with the scaled pion and $\eta$ curves in Fig.~\ref{hades_excess}. This fact provides a hint to the possible origin of the excess yield; pion production at these low energies is known to be dominantly coming from the excitation and decay of baryonic resonances (mainly the $\Delta(1232)$ resonance).

Support for this conjecture comes also from theory~\cite{gale_ca_ca}. Quantitative descriptions of heavy-ion collisions indicate that due to the stopping a high baryon density of up to two times nuclear ground state density is achieved in the center of the collision zone at bombarding energies of $1 - 2$~\gevu. The accompanying temperatures, a quantity not rigourously defined since the thermalization criterium is not necessarily fulfilled when reaching the highest densities, are around $T=80-100$~\mev. One therefore should expect the system not to cross the phase boundary to deconfined matter but rather stay in the hadronic phase throughout. Therefore, the dominant medium radiation should be driven by nucleons and baryonic resonances in the system.

Besides the resonance decays, a strong bremsstrahlung contribution in \np\ interactions has also been predicted within the framework of a covariant OBE\footnote{One Boson Exchange} model~\cite{kaptari}. The quantitative description of these processes in models, however, is difficult. Furthermore, the question whether the observed excess of dileptons is related to any in-medium effect remains open, because of uncertainties in the description of elementary di-electron sources. Precise understanding of the $np \rightarrow npe^{+}e^{-}$ and $pp \rightarrow ppe^{+}e^{-}$ reactions is therefore important for the interpretation of di-electron emission in heavy-ion collisions.

%------------------------------------------------------------------------------
\section{DLS and HADES: "Just a little bit of history repeated?"}
\label{dls_hadeds_relation}
%------------------------------------------------------------------------------
Di-electron production in elementary collisions at $E_{kin}<5$~\gevu\ was also studied by the DLS collaboration. For \pp\ and \pd\ collision systems, complete excitation functions were established~\cite{dls_wilson} and have been available for about $10$ years, although with limited statistics and mass resolution. First theoretical descriptions of di-electron production~\cite{gale_kapusta} stated that $pp$ bremsstrahlung is negligible and that $np$ bremsstrahlung should grow and dominate the di-electron yield as the beam energy increases. This scenario, however, had to be reconsidered after the $\sigma^{p+d} / \sigma^{p+p}$ di-electron yield ratios measured by the DLS experiment became available. The di-electron yield ratio at $4.88$~\gevu\ kinetic beam energy was found to be only two~\cite{dls_ratio_5gev_huang,dls_ratio_5gev_wilson} while it is much higher at lower energies. The theoretical approaches were based on the SPA\footnote{Soft-Photon Approximation} formalism, where only radiation from the external baryon lines is considered and the strong-interaction vertex is treated as on shell ($p^{2} = m^{2}$, $c=1$).

Another approach followed up in Ref.~\cite{kaptari,shyam} is based on an effective OBE model. Unlike the SPA, the OBE formalism allows radiation from internal lines of the interaction diagrams. Drawbacks of the OBE approach include the large number of diagrams which have to be evaluated and ambiguities in adjusting the parameters of the theory. In the OBE calculations, all amplitudes (graphs) are treated coherently. It is important to mention that in the OBE calculations the quantum mechanical interference effects play an important role, essentially reducing the cross section in comparison to an incoherent sum of different contributions. In both cases, \np\ and \pp\ collisions, the interference effects become significant at higher values of the di-electron invariant mass where they reduce the differential cross section by a factor of more than $2$.

However, in the available transport model calculations of dilepton production in $NN$ collisions interference effects are not included. In the transport approach, dileptons from virtual nucleon bremsstrahlung are calculated within the SPA model, restricting the emission process to elastic $NN$ collisions, and the $\Delta$ contribution is treated explicitly by producing and decaying the resonance within a Dalitz-decay model in inelastic collisions. In this way, the interference of elastic and inelastic channels is neglected. The HSD~\cite{transport_hsd} transport model uses a parametrization of virtual bremsstrahlung motivated by a recent OBE~\cite{kaptari} calculation, treating only the elastic channel, since the $\Delta$ decay is treated in the code explicitly. By adjusting the calculations such as to reproduce the cross section for virtual bremsstrahlung as calculated in the OBE approach for elastic channels, the HSD code is now able to interpret the \pp\ and \xdp\ data measured by the DLS collaboration, and also, can describe the heavy-ion data of both, the DLS and HADES experiments. The solution of the "DLS puzzle" is hence claimed by the authors~\cite{bratkovskaya}.

In the following, the results obtained by HADES for di-electron production in elementary collisions will be used to check this conclusion. But, in contrast to the DLS, who recorded events with proton beam on a deuterium target, HADES used \xdp\ reactions and selected "quasi-free" \np\ interactions via detection of the spectator protons in the forward spectrometer.

%==============================================================================
\section{The HADES strategy}
%==============================================================================
As it has been discussed before, the major experimental challenge is to discriminate the penetrating but very rare leptons from the huge hadronic background which exceeds the electron signal by many orders of magnitude. The HADES~\cite{hades_tech} detector has been specifically designed to overcome these difficulties. It is set up at SIS18 to study di-electron production in heavy-ion as well as elementary and pion-induced reactions in the energy regime of $1-2$~\gevu\ in a systematic way, with high quality data.

To contribute to a better understanding of the contributions to di-electron production in the early stage of heavy-ion collisions, HADES has studied \pp\ and \xdp\ interactions at $E_{kin} = 1.25$ \gevu, i.e.\ below the $\eta$ meson production threshold in proton-proton reactions. The main goal of the latter experiment is to understand the \np\ bremsstrahlung component for \ee\ production in the tagged reaction channel $np \rightarrow npe^{+}e^{-}$ and to establish an experimental cocktail of di-electrons from "free", i.e.\ non-medium hadron decays for SIS energies. A direct separation of the virtual bremsstrahlung and $\Delta$ Dalitz decays is not possible, due to the interference effects. I addition, electron pairs show similar spectral and transverse momentum distribution in both processes. However, important information can be obtained from the di-electron yield observed in \pp\ and \np\ reactions, since they present a different sensitivity to the virtual bremsstrahlung and $\Delta$ Dalitz decays.

Our investigation of electron pair production in elementary reactions of protons and neutrons on a proton target started with a \pp\ experiment at $1.25$~\gev\ kinetic energy. A beam of up to $10^{7}$ protons per second was incident on a $5$~cm liquid hydrogen target. In total about $770$~M electron enriched events were taken during a period of $9$ days in April~$2006$. The program was continued with a two-week beam time in May~$2007$ when neutron-induced reactions were measured using a deuterium beam of $1.25$~\gevu. Here, the \pp\ reactions can be suppressed by measuring the spectator proton in a Forward hodoscope Wall (FW) covering the polar angle region between $1^{\circ}$ and $7^{\circ}$. The \ee\ pairs were detected in coincidence with the fast spectator protons. From the reconstructed like-sign invariant-mass distributions, the respective combinatorial background distribution was obtained as the arithmetical mean of like-sign $e^{+}e^{+}$ and $e^{-}e^{-}$ pairs. The di-electron yield measured in HADES was furthermore corrected for detection and reconstruction inefficiencies. Figs.~\ref{pp_hsd_iqmd} and~\ref{dp_hsd_iqmd} show the resulting \ee\ invariant-mass distributions of true pairs normalized to the number of pp elastic scattering measured in HADES (for \pp\ and \np\ runs respectively) and extrapolated to the full solid angle. The $\pi^{0}$ cross sections used for normalization of the experimental data were taken from Ref.~\cite{cross_section}. The total systematic errors of $\sim30\%$ (shown as horizontal error bars in Figs.~\ref{pp_hsd_iqmd},~\ref{dp_hsd_iqmd}) are dominated by uncertainties caused by the electron efficiency correction and uncertainties in the $pp$ elastic scattering normalization and the extrapolation procedure.

%==============================================================================
\section{$p$+$p$ and $n$+$p$ data in transport model calculations}
%==============================================================================
%
\begin{figure}[!th]
\centering
\mbox{{\includegraphics[width=0.5\textwidth]{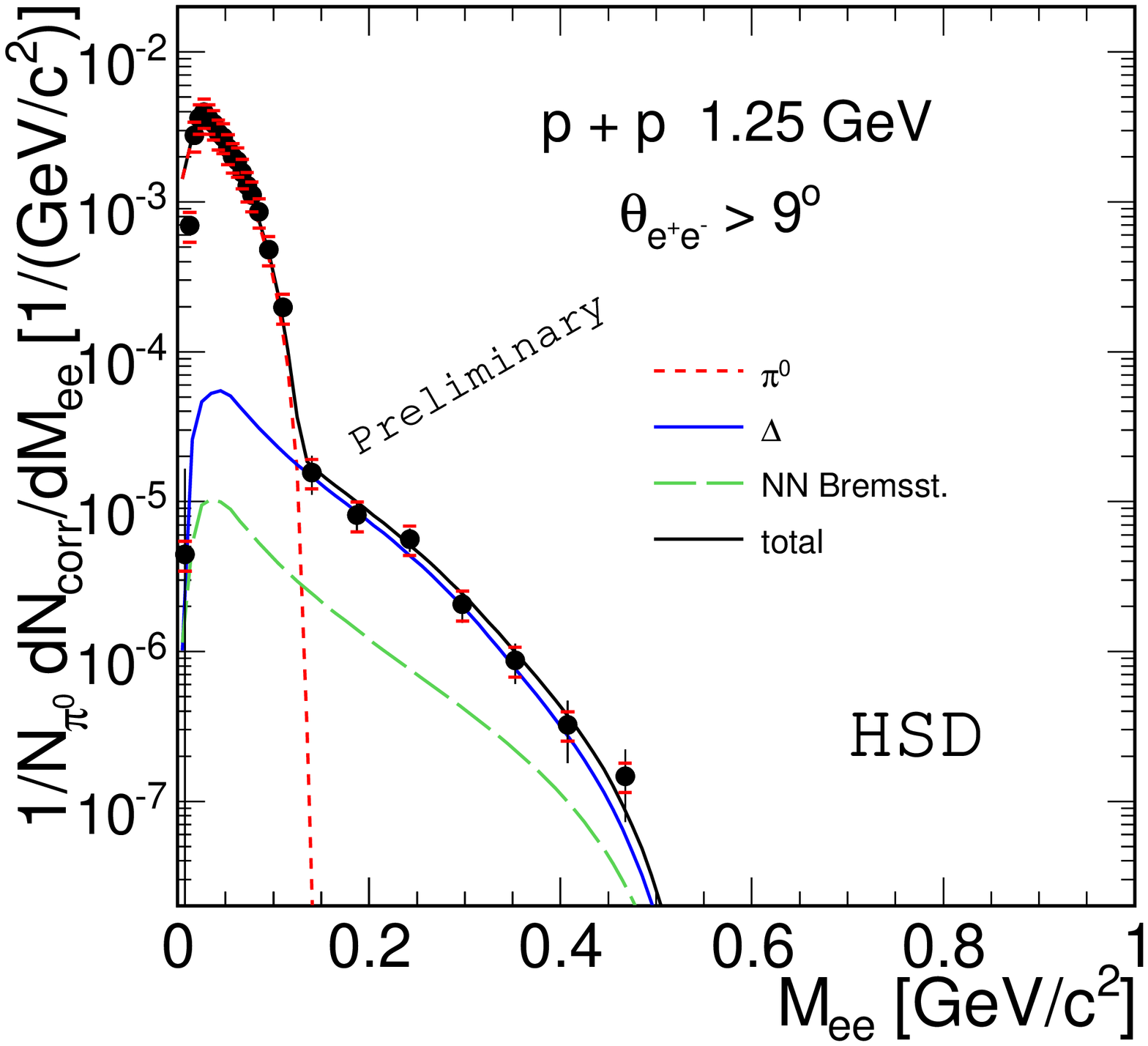}}
    \quad
    {\includegraphics[width=0.5\textwidth]{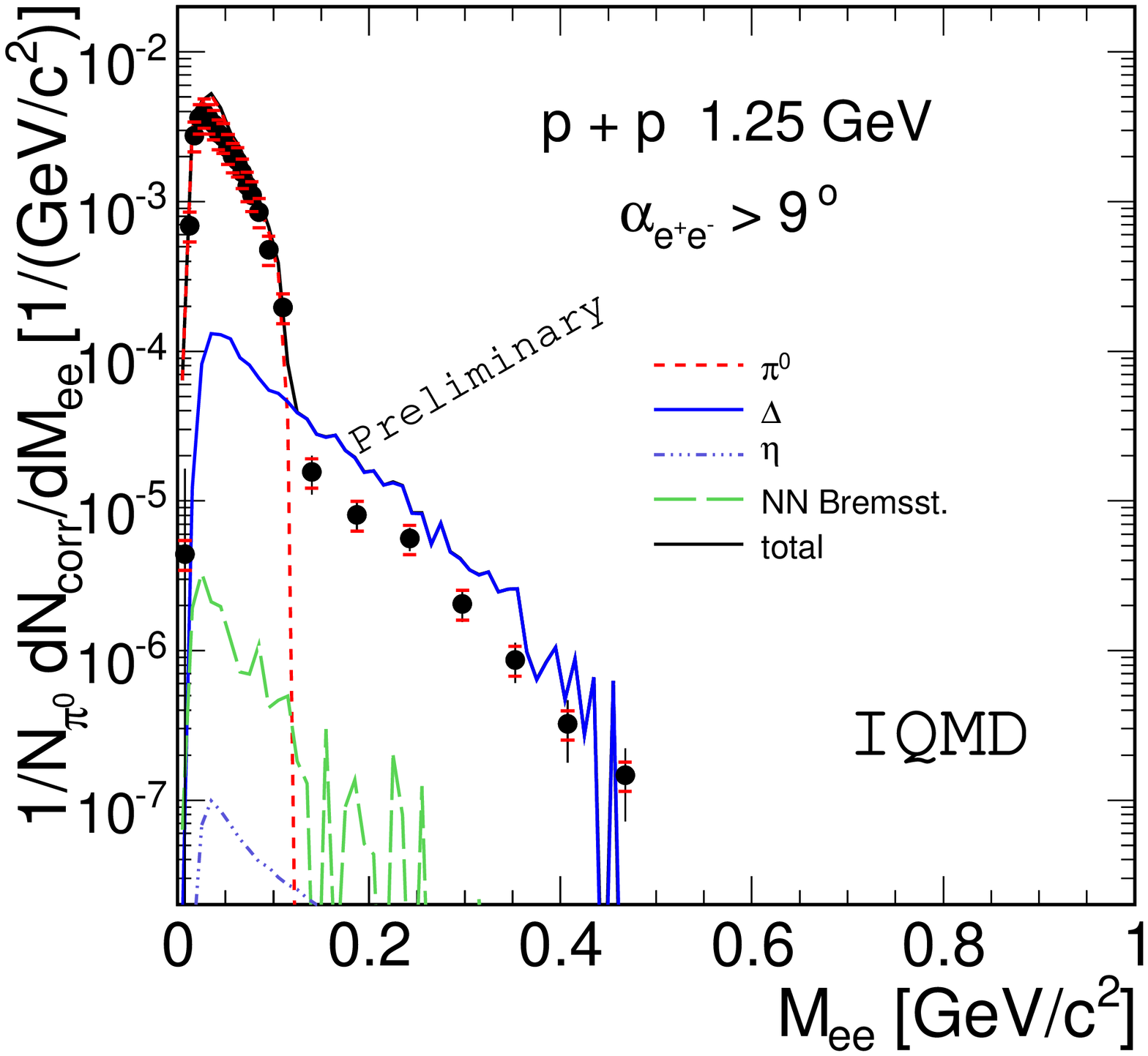}}}
        \caption{Invariant-mass distribution of \ee\ pairs measured in \pp\ interactions at a beam energy of 1.25~\gev\ compared with HSD (left panel) and IQMD (right panel) transport model calculations. Dashed-dotted line: $\pi^{0}$~Dalitz, red solid line: $\Delta^{+}$~Dalitz, green dashed line: $NN$ bremsstrahlung, black solid line: total cocktail.}
\label{pp_hsd_iqmd}
\end{figure}
\begin{figure}[!h]
\centering
\mbox{{\includegraphics[width=0.5\textwidth]{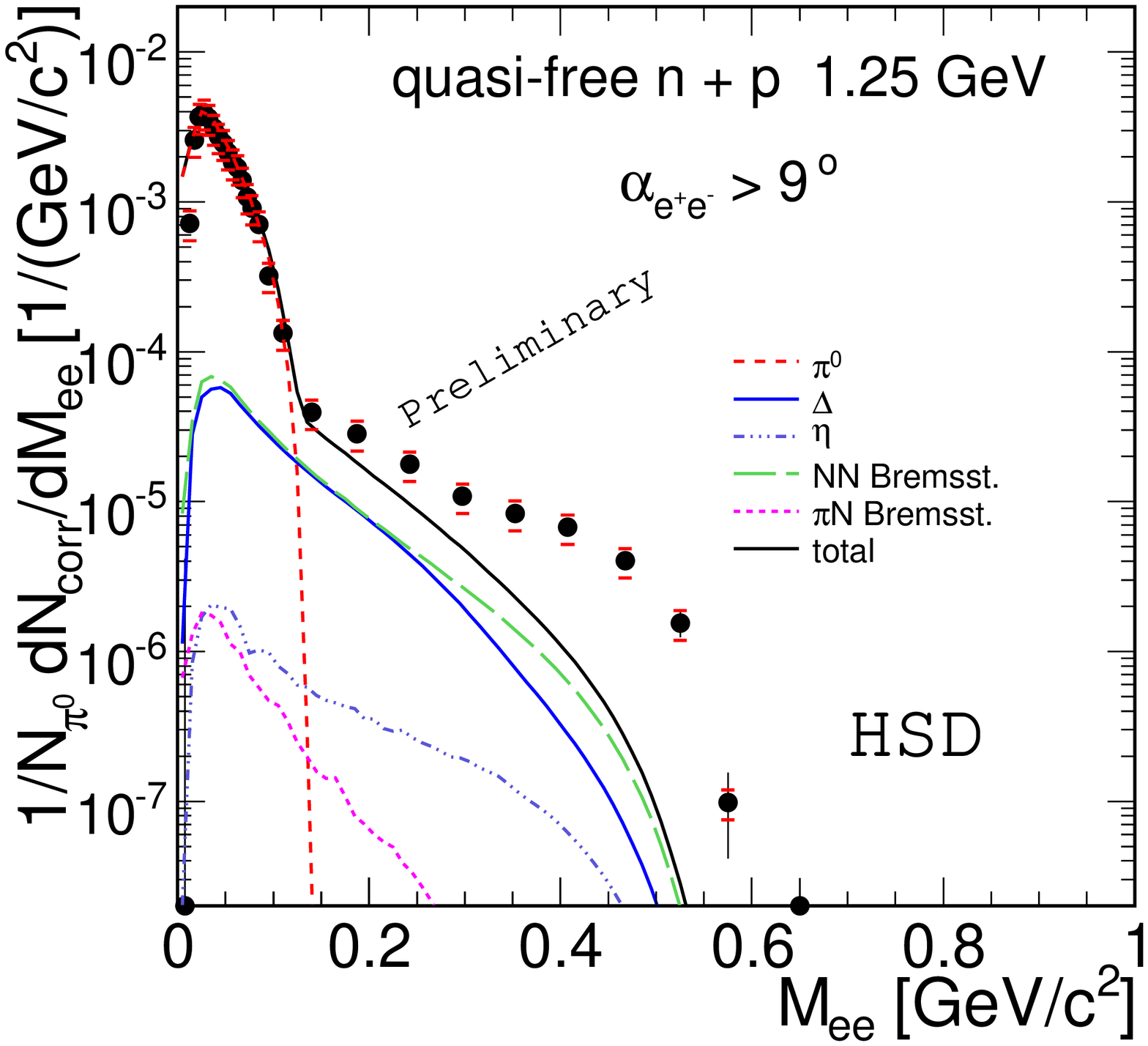}}
    \quad
        {\includegraphics[width=0.5\textwidth]{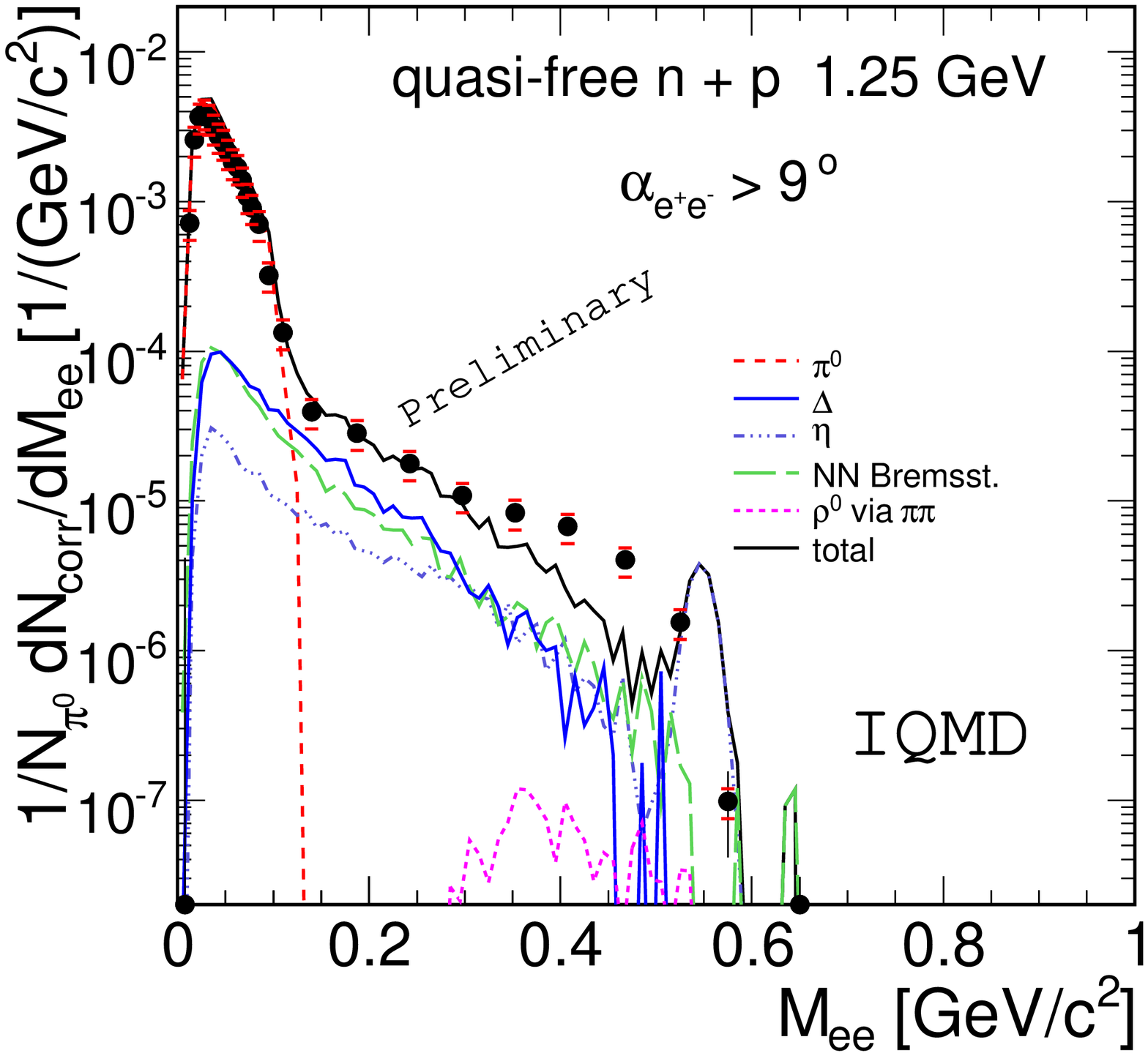}}}
            \caption{Invariant-mass distribution of \ee\ pairs measured in \xdp\ interactions at a beam energy of 1.25~\gevu\ compared with HSD (left panel) and IQMD (right panel) transport model calculations.
            Dashed-dotted line: $\pi^{0}$~Dalitz, red solid line: $\Delta^{+,0}$ Dalitz, green dashed line: $NN$ bremsstrahlung, blue dashed line: $\eta$. Dashed magenta line: HSD - $\pi N$ bremsstrahlung, IQMD - $\rho$
            meson production via $\pi\pi$ annihilation. Black solid line: total cocktail.}
\label{dp_hsd_iqmd}
\end{figure}
Obviously, it is not a trivial task to interpret the di-electron data at SIS/Bevalac energies. Transport models have to incorporate all the various sources that contribute to the experimental spectra. In order to compare our results to theoretical descriptions, di-electrons generated with the HSD\footnote{HSD version from October $2007$ including $NN$ bremsstrahlung \`{a}-la Kaptari~\etal\ is used here} and the IQMD~\cite{iqmd} transport models, respectively, were filtered with the HADES acceptance and normalized to the respective $\pi^{0}$ multiplicity. Figure~\ref{pp_hsd_iqmd} shows the inclusive di-electron invariant mass distribution obtained in \pp\ collisions at a beam energy of $E_{kin} = 1.25$~\gev. The data are compared with the HSD transport calculations performed by Bratkovskaya \etal~\cite{bratkovskaya} (see Fig.~\ref{pp_hsd_iqmd}, left panel) and results of IQMD transport calculations performed by Aichelin~\etal\ \cite{transport_qmd} (see Fig.~\ref{pp_hsd_iqmd}, right panel). At low masses, a prominent $\pi^{0} \rightarrow e^{+}e^{-}\gamma$ peak dominates the spectra. In the intermediate-mass region Dalitz decays of $\Delta \rightarrow Ne^{+}e^{-}$ and $NN$ bremsstrahlung are the major sources.

The HSD model reproduces the \pp\ experimental data in the whole mass range quite well. On the other hand, the IQMD model, still using certain previous parameterizations of the bremsstrahlung, can reproduce the shape of the \pp\ mass spectra, however, has a problem with the overall normalization on the $40\%$ level in the whole mass range. It is also seen that the relative contributions of individual components like from $\Delta$ Dalitz-decay and $NN$ bremsstrahlung are very different in the two models. The situation becomes more difficult with the \np\ data. Figure~\ref{dp_hsd_iqmd} shows inclusive di-electron invariant-mass distributions obtained in quasi-free \np\ collisions at a beam energy $E_{kin} = 1.25$~\gev\ together with the transport model calculations. One should point out that the experimental spectator proton measurement is not fully implemented in these calculations. Nevertheless, this first comparison shows that both models fail in reproducing the quasi-free \np\ data in the mass region above the $\pi^{0}$ Dalitz-decay range. Again, the contribution of individual sources differs by factors between the models. The quasi-free \np\ data clearly shows beyond errors that something is still missing in the transport calculations. Further, normalizing the efficiency corrected $\frac{1}{2}(pp+pn)$ spectrum to $\pi^0$ and comparing it to the \xcc\ invariant-mass spectrum measured at $1$~\gevu\ beam energy (also normalized to $\pi^0$) shows that the superposition of elementary collisions is sufficient to describe the observed di-electron yield in \xcc\ collisions.

%==============================================================================
\section{R\'{e}sum\'{e} and prospects}
%==============================================================================
Our preliminary data for the invariant-mass spectrum in the reaction $np \rightarrow  np e^+e^-$, extracted from the tagged subreaction in $dp \rightarrow p_{spectator}np e^+e^-$, point to a shoulder at intermediate values of the di-electron invariant mass. Such a structure is hardly described within the transport approach. As it has been shown, understanding the elementary channels remains challenging.

An investigation of the medium-heavy \arkcl\ system recently investigated shows a pair excess beyond the one observed in \xcc\ reactions. Furthermore, the di-electron invariant-mass distribution shows for the first time a clear $\omega$ signal at SIS energies. The presently available HSD and UrQMD transport code predictions overestimate the $\rho^{0}$ production~\cite{filip_qm09}.

The $\omega$ meson has furthermore been produced and identified in dedicated experiments, measuring in $2007$ the \pp\ and in $2008$ the $p$+$Nb$ reactions at $3.5$~\gev. These combined measurements provide the means to address directly in-medium effects on vector mesons in cold nuclear matter.

The investigation of di-electron production in heavy-ion collisions within the FAIR project is also planned. The ultimate goal is to provide a complete excitation function of dilepton production up to energies of $45$~\gevu: with HADES covering energies up to $8$~\gevu\ and with Compressed Baryonic Matter (CBM) experiment taking over for higher beam energies~\cite{stroth_hades_cbm}.

%==============================================================================

%==============================================================================

\end{document}